\documentclass[11pt,twoside]{article}
\usepackage{asp2010}

\resetcounters

\begin{document}

\title{Observations and Modelling of DQ White Dwarfs}
\author{Tommi Vornanen,$^1$ Svetlana Berdyugina,$^2$ and Andrei Berdyugin$^1$
\affil{$^1$Department of Physics and Astronomy, University of Turku, Turku, Finland}
\affil{$^2$Kiepenheuer Institute f\"ur Sonnenphysik, Freiburg, Germany}}

\begin{abstract}
We present spectropolarimetric observations and modelling of 12 DQ white dwarfs.
Modelling is based on the method presented in \citet{ber05}. We use the
model to fit the C2 absorption bands to get atmospheric parameters in different
configurations, including stellar spots and stratified atmospheres, searching for the
best possible fit. We still have problem to solve before we can give temperature estimates
based on the Swan bands alone.
\end{abstract}

\section{Introduction}
We have observed 12 DQ white dwarfs to
search for magnetic members in the class.
Many of the few known peculiar DQs
\citep{sch99} and hot DQs \citep{duf07} are magnetic but the more
numerous subclass, normal DQs had, until
recently, only one magnetic member, G99-
37. From our survey, we discovered that
GJ841B is also magnetic and its properties
closely resemble those of G99-37 \citep{vor10}.

Of the 12 stars GJ841B is the only clearly
magnetic case. WD1235+422 also has a
polarized spectrum but we haven't been able
to model it with enough detail to say
anything more about it.

\section{Method}
Figure \ref{fig1} shows the intensity spectrum of
GJ893.1 and the best fitting model that we
could originally find.
\articlefigure[width=8 cm]{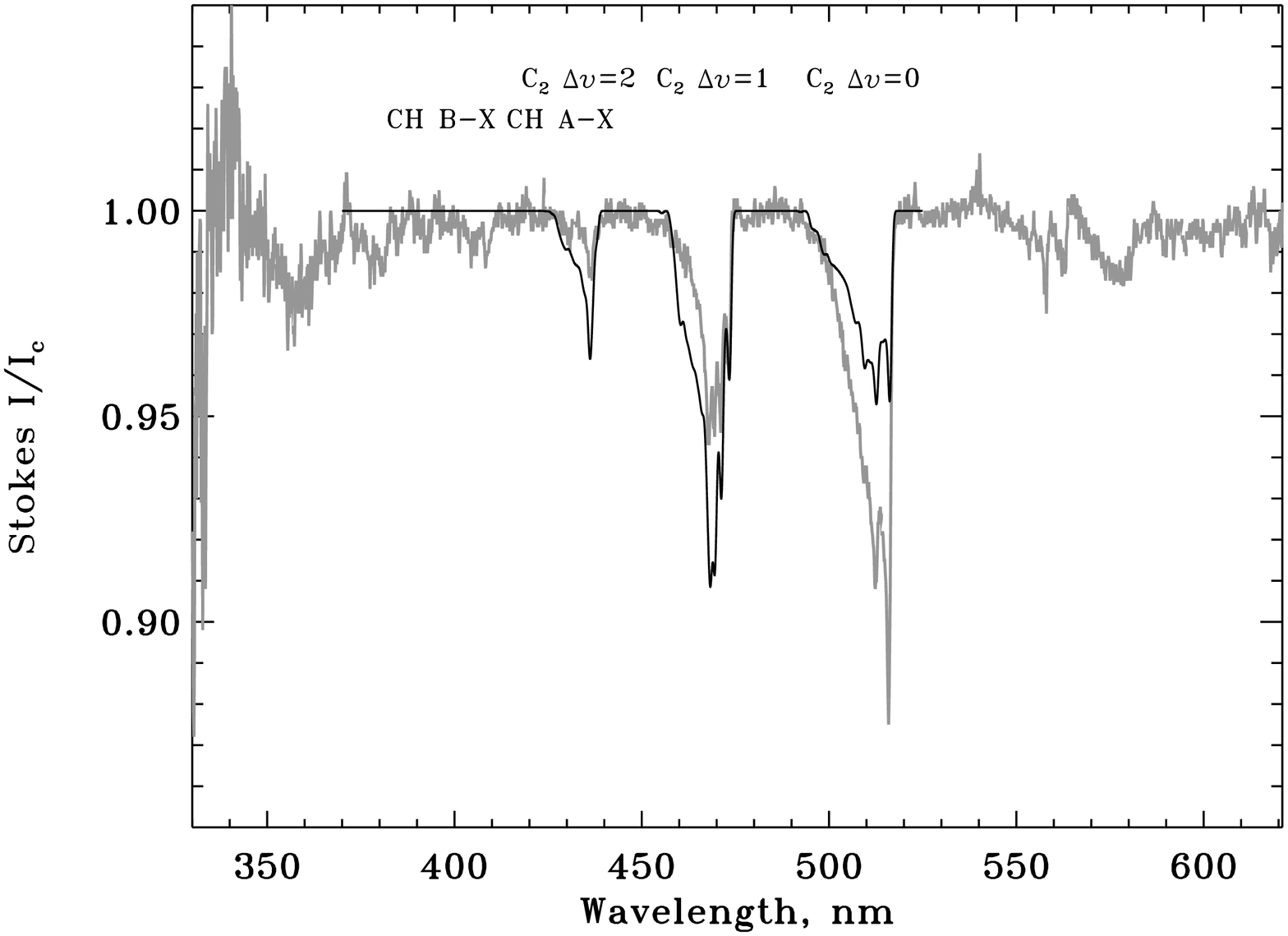}{fig1}{GJ893.1 first modelling.}
It is clear that the different molecular bands
have wrong relative depths. However, the
model has worked very well for CH features
in G99-37 and GJ841B so it should work for
C2 too. To improve the fit, we tried other
physical configurations.

First, we explored a stellar spot. Figure \ref{fig2}
shows the resulting spectrum for a
reasonable combination of spot and
atmospheric temperatures.
\articlefigure[width=8 cm]{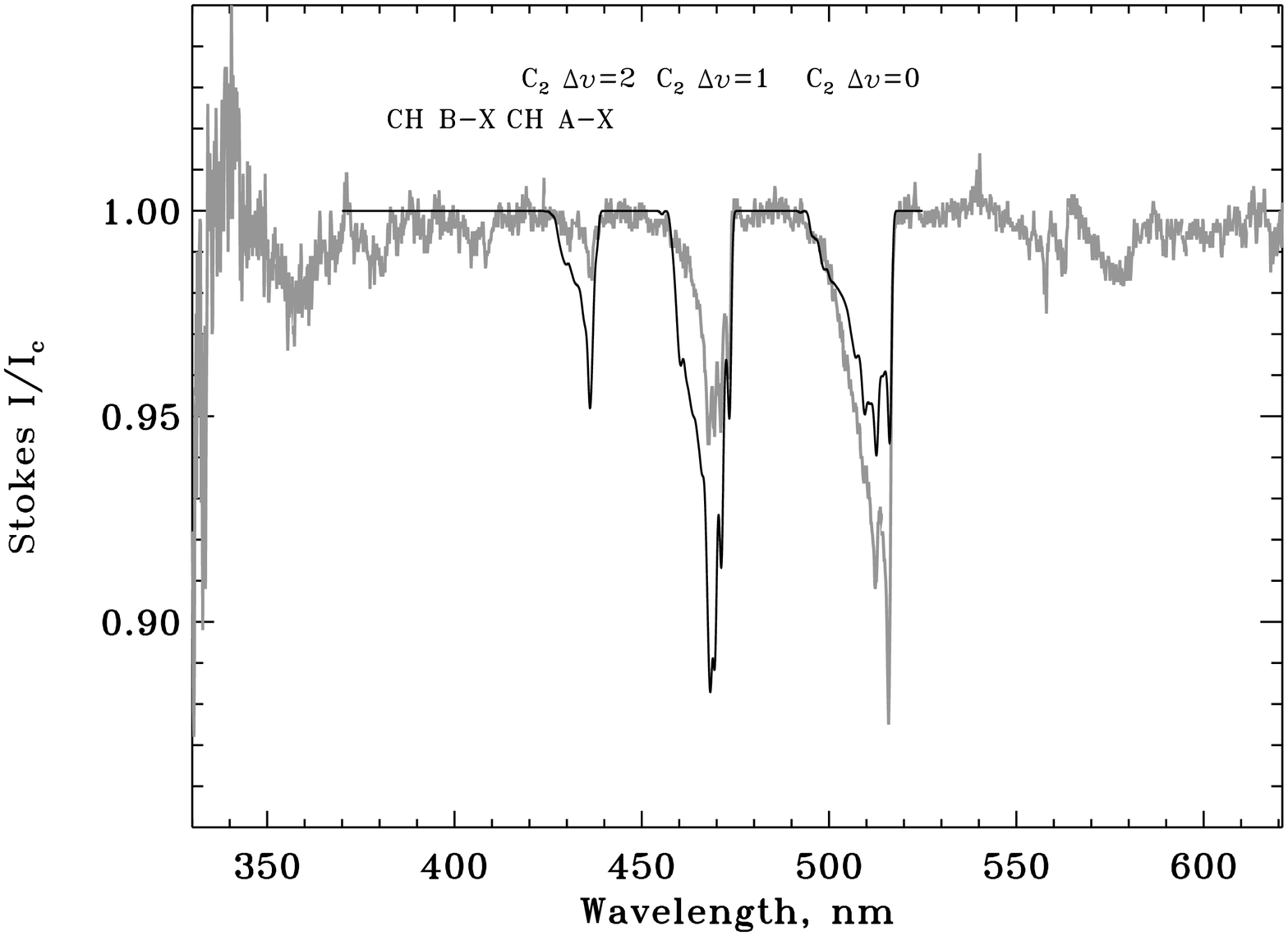}{fig2}{Spot model for GJ893.1.}
It is easy to see that a spot model does not
improve the fit unless one uses different spot
sizes for different absorption bands, which is
unrealistic.

Next, we tried a stratified atmosphere.
We allowed for three different layers, each
with a different temperature and carbon
abundance. The resulting fit is already a little
bit better (See Fig. \ref{fig3}) but the relative depths
of the Swan bands still differ.
\articlefigure[width=8 cm]{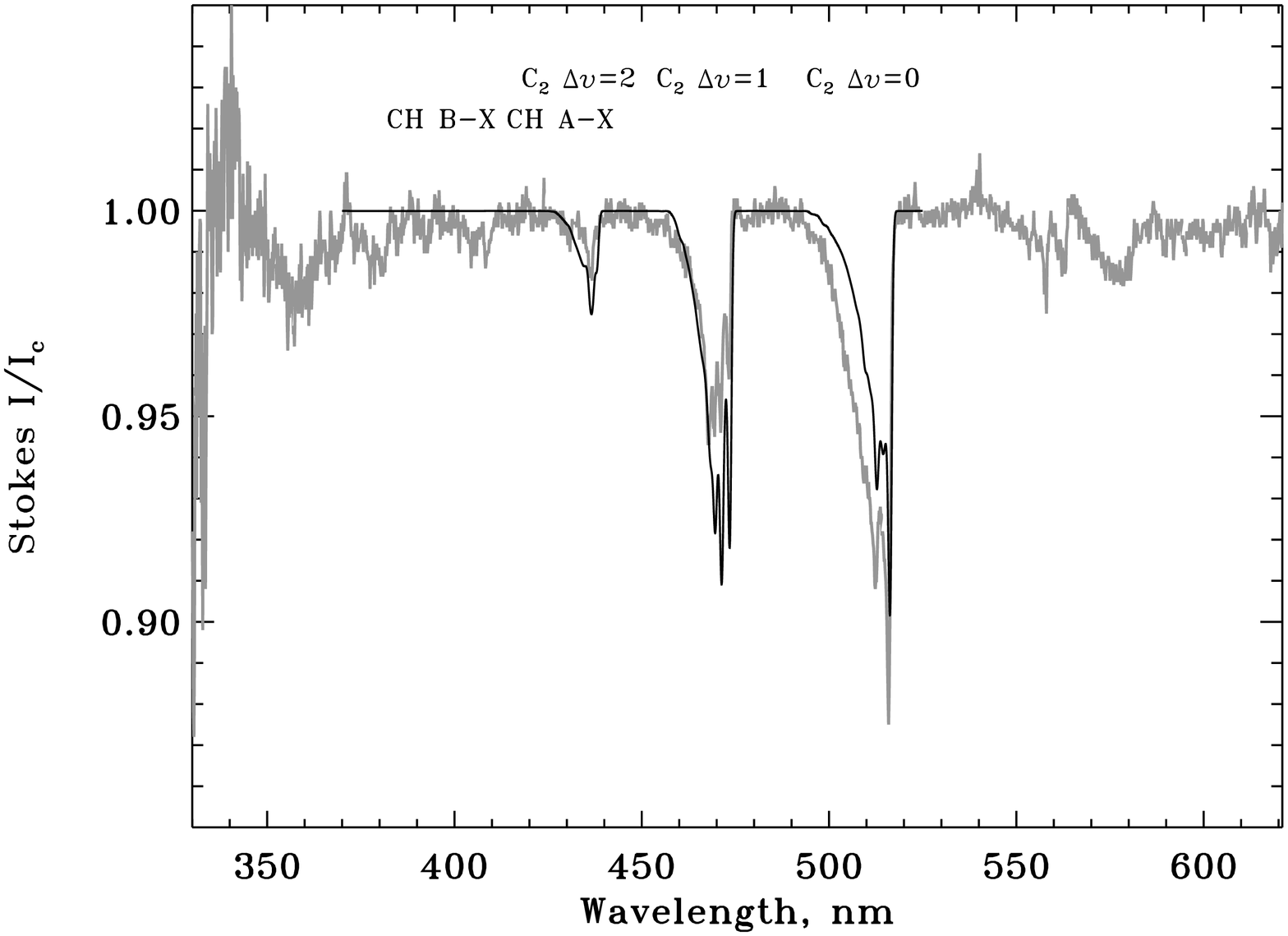}{fig3}{Stratified atmosphere with three layers.}
The best fitting model combinations are, however,
unrealistic. Temperatures of 2000-3000 K
are favored and this is too low for a white
dwarf according to the current
understanding.

Finally, we decided to modify the
oscillator strengths of C2 to get a better fit.
These parameters control the amount of
absorption from each transition in the
molecule. The values are determined in a
laboratory, so changing them is a bit
dubious, but the resulting fit is very good
compared to the previous ones (Fig. \ref{fig4}). We also increased the maximum value of the total angular momentum quantum number J for the C$_2$ molecule. This is responsible for the better fits in the blue wing of the absorption bands.
\articlefigure[width=8 cm]{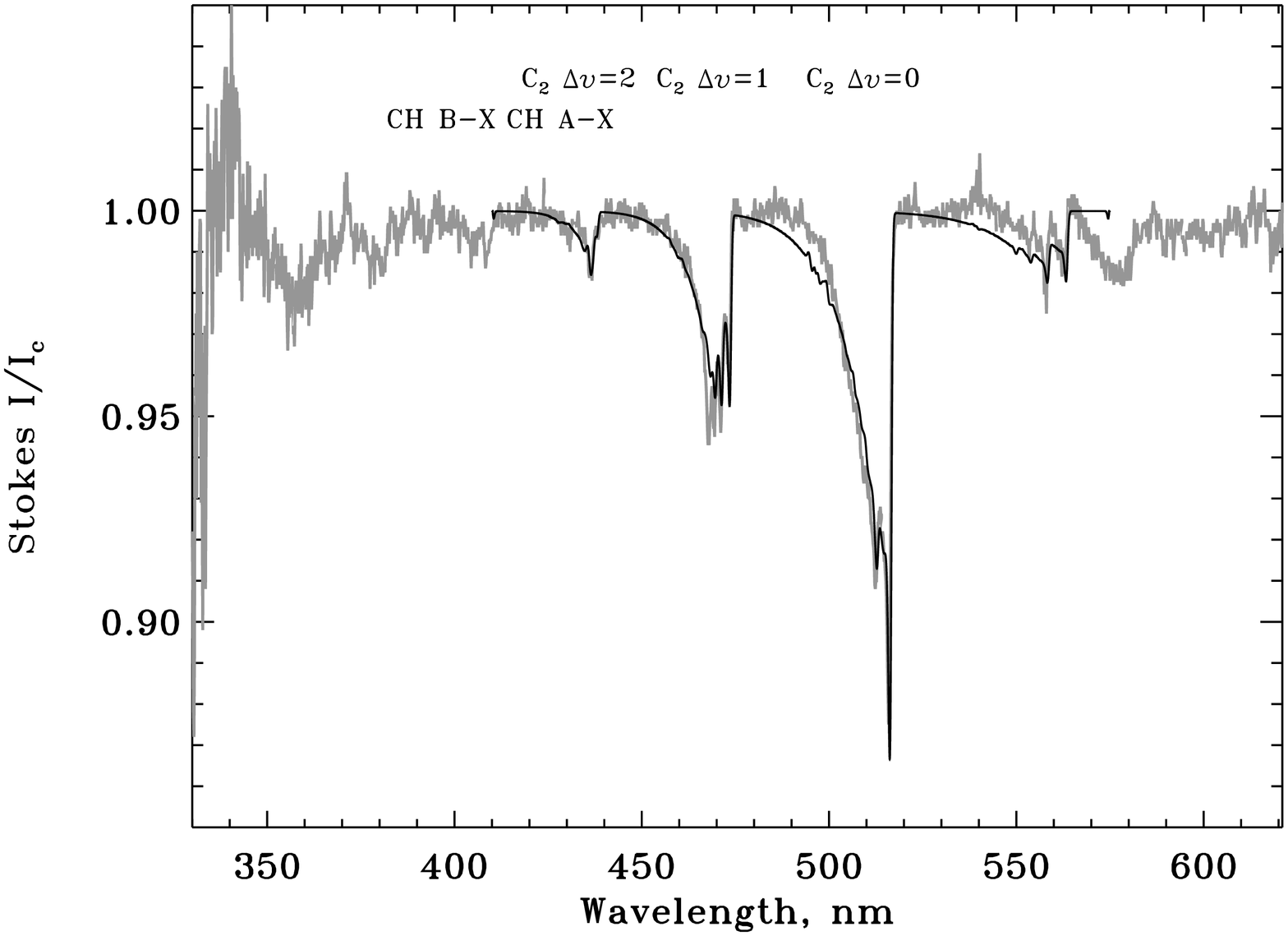}{fig4}{Final fit for GJ893.1 with modified
oscillator strengths.}

\section{Results}
Figure \ref{fig5} shows model fits for the non-magnetic
DQ WDs in our survey as well as
WD1235+422, which might be magnetic.
Atomic lines in GJ1117 affect the automated
fitting process and the best fit is actually
very poor for that star.
\articlefigure[width=8 cm]{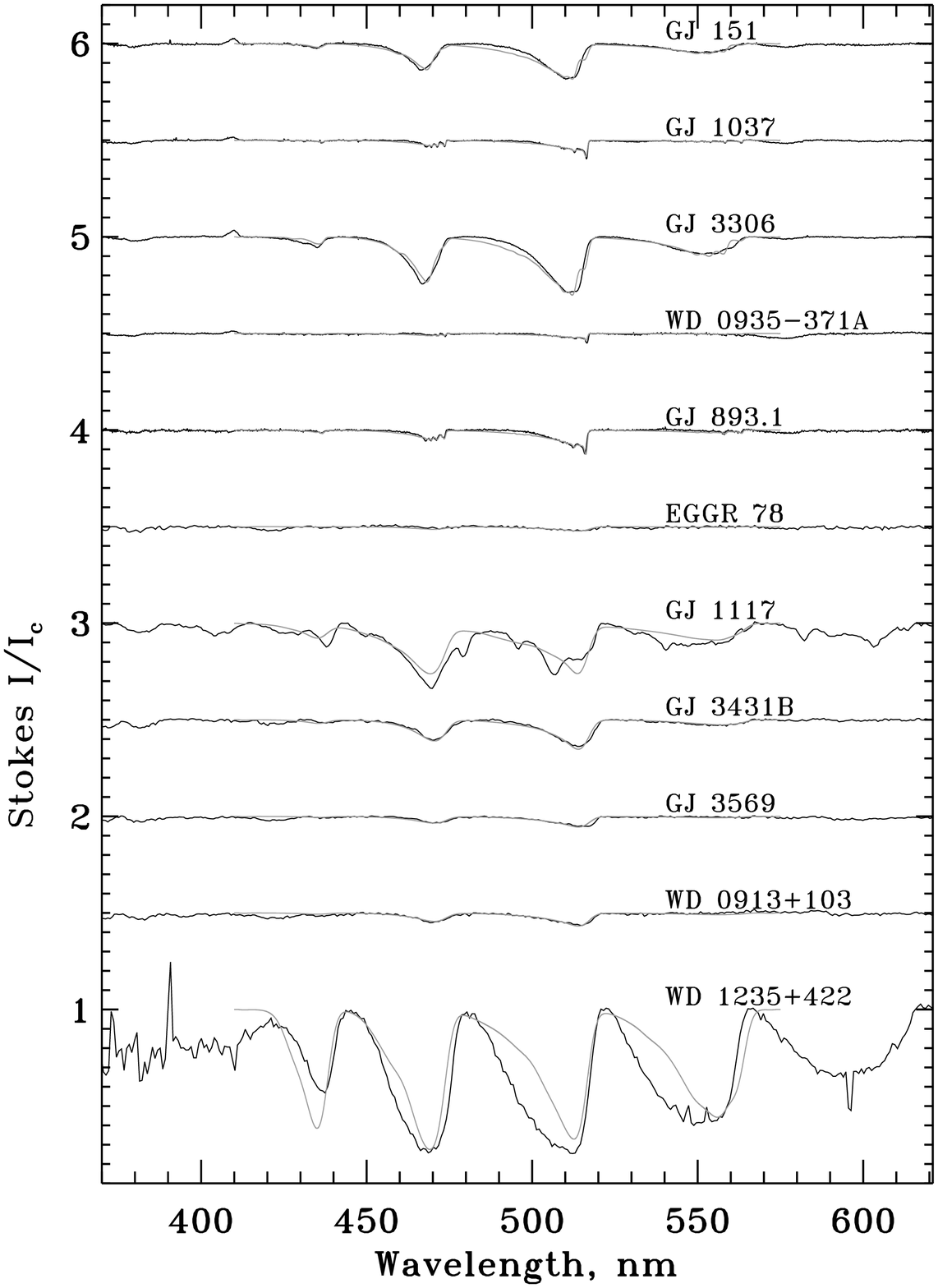}{fig5}{Observed spectra and best fitting models
for our survey targets excluding GJ841B.}

\section{Obstacle}
One problem remains, however. There is a
free parameter in the code that controls the
strengths of the individual absorption lines.
This number is used to multiply the line
profile to modify its depth. The parameter
must be connected with a physical
parameter (probably carbon abundance)
and once that connection is found, we can
provide temperature estimates for these
white dwarfs based on our model.

\acknowledgements T.V. would like to thank the Finnish Graduate School in Astronomy and Space Science for funding during this research.

\bibliography{vornanen}

\end{document}